\documentclass[groupedaddress,aps,pra,superscriptaddress,showpacs,twocolumn,prl]{revtex4}%
\usepackage{epsfig,dsfont,amssymb,amsmath,amsthm,amsfonts,amsbsy,mathrsfs}
\usepackage{graphicx, color}
\usepackage{epstopdf}
\usepackage{mathdots}
\usepackage{float}
\usepackage{xcolor}
\DeclareMathOperator{\Tr}{Tr}
\begin{document}

\title{Concurrence Triangle Induced Genuine Multipartite Entanglement Measure}

\author{Zhi-Xiang Jin}
\affiliation{School of Computer Science and Technology, Dongguan University of Technology, Dongguan, 523808, China}
\affiliation{School of Physics, University of Chinese Academy of Sciences, Yuquan Road 19A, Beijing 100049, China}
\affiliation{Max-Planck-Institute for Mathematics in the Sciences, Leipzig 04103, Germany}
\author{Yuan-Hong Tao}
\affiliation{School of Science, Zhejiang University of Science and Technology, 318 Liuhe Road, Hangzhou,
Zhejiang 310023, China}
\author{Yao-Ting Gui}
\affiliation{Max-Planck-Institute for Mathematics in the Sciences, Leipzig 04103, Germany}
\author{Shao-Ming Fei}
\thanks{Corresponding author: feishm@cnu.edu.cn}
\affiliation{Max-Planck-Institute for Mathematics in the Sciences, Leipzig 04103, Germany}
\affiliation{School of Mathematical Sciences,  Capital Normal University,  Beijing 100048,  China}

\author{Xianqing Li-Jost}
\affiliation{Max-Planck-Institute for Mathematics in the Sciences, Leipzig 04103, Germany}

\author{Cong-Feng Qiao}
\thanks{Corresponding author: qiaocf@ucas.ac.cn}
\affiliation{School of Physics, University of Chinese Academy of Sciences, Yuquan Road 19A, Beijing 100049, China}
\affiliation{CAS Center for Excellence in Particle Physics, Beijing 100049, China\\ \vspace{7pt}}
\bigskip

\begin{abstract}
We study the quantification of genuine multipartite entanglement (GME) for general multipartite states. A set of inequalities satisfied by the entanglement of $N$-partite pure states is derived by exploiting the restrictions on entanglement distributions, showing that the bipartite entanglement between each part and its remaining ones cannot exceed the sum of the other partners with their remaining ones. Then a series of triangles, named concurrence triangles, are established corresponding to these inequalities. Proper genuine multipartite entanglement measures are thus constructed by using the geometric mean area of these concurrence triangles, which are non-increasing under local operation and classical communication. The GME measures classify which parts are separable or entangled with the rest ones for non genuine entangled pure states. The GME measures for mixed states are given via the convex roof construction, and a witness to detect the GME of multipartite mixed states is presented by an approach based on state purifications. Detailed examples are given to illustrate the effectiveness of our GME measures.
\end{abstract}

\maketitle

\section{INTRODUCTION}
As an important resource multipartite entanglement plays an significant role in quantum communication and quantum information processing. Although the experimental observations of multipartite entanglement have been successfully implemented \cite{YPY,NFO,VSA}, its rigorous characterization is far from being satisfied. Different from bipartite entanglement, for multipartite systems one has so called ``genuine multipartite entanglement" (GME) \cite{SHW,GVP,MAA}. A key issue is to give a suitable measure of GME for quantifying the genuine multipartite entanglement.

In three-qubit systems, the GHZ class and the W class states are both GME ones that are neither the product states nor biseparable states \cite{WDG}. The GHZ state is more entangled than the W state in the sense that the GHZ state can be used to faithfully teleport an arbitrary single-qubit quantum state, while the W state is relatively less capable \cite{joo}. In the other sence, however, GHZ state is not more entangled than W state, such as 1-1 tangle negativity in the noninertial frame \cite{dq1,dq2}. There are three known GME measures for three-qubit systems, which are either equivalent or dependent. The first GME measure, genuinely multipartite concurrence (GMC), presented by Ma et al. \cite{mzh} and further developed by Hashemi Rafsanjani et al. \cite{smh}, is exactly the minimum concurrence between each single qubit and its remaining partners. The second GME measure, generalized geometric measure, is given by Sen(De) and Sen \cite{us1,us2}, which is based on the distance between a given state and its closest biseparable states. The third GME measure is proposed by Emary and Beenakker \cite{cw}, which is actually the average of 3-tangle and GMC in \cite{mzh}.

%In \cite{guoy}, the authors give an approach of constituting genuine $m$-partite entanglement measures by using the product of the radius of inscribed circle and circumscribed circle of a triangle.

A well defined GME measure has to satisfy the following conditions. (a) The measure must be zero for all product and biseparable states. (b) The measure must be positive for all non-biseparable states. (c) The measure should be nonincreasing under local operations and classical communications (LOCC).

%Xie and Eberly \cite{xsb} believe that a proper GME measure, specially in three-qubit systems, should add a new condition: (c) The measure ranks the GHZ state as more entangled than the W state.

Very recently, Xie and Eberly \cite{xsb} defined a new GME measure particularly for three-qubit systems, which has a simple form and an elegant geometric interpretation, together with superiorities to the above three known GME measures. From this measure it is verified that the GHZ state is more entangled than W state. Unfortunately, it has been shown that this measure is increasing under local operations and classical communications (LOCC)\cite{csm}, which means it is not a proper entanglement measure.

In this work, we define proper GME measures for general multipartite qudit systems. By considering the restrictions among all the entanglement between a single qudit and the remaining ones in a multipartite system, we establish a set of polygamy inequalities in terms of (squared) concurrence, showing that the bipartite entanglement between each part and its remaining ones cannot exceed the sum of the other partners with their remaining. We then illustrate that these inequalities can be regarded as a set of concurrence triangles, i.e., the three one-to-other concurrences can represent the lengths of the three edges of a triangle. By using these polygamy inequalities, we advance a GME measure for multipartite states based on the geometric mean area of the concurrence triangles, which satisfies all the three requirements (a), (b) and (c) of a bona fide GME measure.

\section{ Polygamy Inequalities and Their Geometric Implications}
Let $H_X$ denote a $d$-dimensional vector space associated with the system $X$. For a bipartite pure state $|\psi\rangle_{AB}$ in vector space $H_A\otimes H_B$, the concurrence is given by \cite{AU,PR,SA},
$C(|\psi\rangle_{AB})=\sqrt{{2\left[1-\mathrm{Tr}(\rho_A^2)\right]}}$,
where $\rho_A=\mathrm{Tr}_B(|\psi\rangle_{AB}\langle\psi|)$ is the reduced density matrix by tracing over the subsystem $B$. Let $T(\rho)$ denote the linear entropy of state $\rho$, $T(\rho)=1-\mathrm{Tr}(\rho^2)$ \cite{EM}. For a bipartite state $\rho_{AB}$, one has \cite{CYY},
\begin{eqnarray}\label{LS}
|T(\rho_A)-T(\rho_B)|\leq T(\rho_{AB})\leq T(\rho_A)+T(\rho_B).
\end{eqnarray}

In the following, we consider general $N$-qudit systems with subsystems $A_1,...,A_N$. For simplicity we denote the concurrence between the subsystem $A_i$ and the rest subsystems of a pure state $|\psi\rangle\in H_{A_1}\otimes ...\otimes H_{A_N}$ as
$C_{i|\widehat{i}}(|\psi\rangle):=C_{A_i|A_1...A_{i-1}A_{i+1}...A_N}(|\psi\rangle)$, where $\widehat{i}=1\cdots(i-1)(i+1)\cdots N$ stands for subsystem $A_1...A_{i-1}A_{i+1}...A_N$, i.e., $\hat {i}$ stands for $i$ being omitted in the subindices.

{\bf Theorem 1.} For any $N$-partite pure state $|\psi\rangle\in H_{A_1}\otimes ...\otimes H_{A_N}$, we have
\begin{eqnarray}\label{th1}
C^2_{i|\widehat{i}}(|\psi\rangle)\leq\sum_{j\ne i}^NC^2_{j|\widehat{j}}(|\psi\rangle),
\end{eqnarray}
and
\begin{eqnarray}\label{lemma}
C_{i|\widehat{i}}(|\psi\rangle)\leq\sum_{j\ne i}^NC_{j|\widehat{j}}(|\psi\rangle).
\end{eqnarray}

{\sf Proof.} For any $N$-partite pure state $|\psi\rangle$, we have
\begin{eqnarray*}
C^2_{i|\widehat{i}}(|\psi\rangle)&&=2\left(1-\Tr(\rho^2_{ {\widehat{i}}})\right)=2T(\rho_{ \widehat{i}})\\
&&\leq \sum_{j\ne i}2T(\rho_{j})=\sum_{j\ne i}^NC^2_{j|\widehat{j}}(|\psi\rangle),
\end{eqnarray*}
where the inequality is due to (\ref{LS}).

The inequality (\ref{lemma}) is easily deduced by (\ref{th1}) since
$\sum_{j\ne i}^NC^2_{j|\widehat{j}}(|\psi\rangle)\leq(\sum_{j\ne i}^NC_{j|\widehat{j}}(|\psi\rangle))^2$. $\Box$

These inequalities in Theorem 1 are valid for any $N$-partite pure state $|\psi\rangle$ which include the results of \cite{qxf} as special cases of $N$-qubit pure states.
Obviously, these polygamy inequalities guarantee that all the (squared) one-to-rest qudit concurrences, representing the lengths of edges, form a closed $N$-sided polygon. We may also interpret them as the lengths of edges for a series of triangles. We name them as the concurrence triangles.

From the inequality (\ref{lemma}), taking into account the bipartition $ij|\widehat{ij}$, we have
\begin{eqnarray}\label{4p}
C_{\widehat{ij}|ij}(|\psi\rangle)\leq C_{i| \widehat{i}}(|\psi\rangle)+C_{j| \widehat{j}}(|\psi\rangle),\nonumber\\
C_{i| \widehat{i}}(|\psi\rangle)\leq C_{\widehat{ij}|ij}(|\psi\rangle) +C_{j| \widehat{j}}(|\psi\rangle).
\end{eqnarray}
Here, $\{i,j,\widehat{ij}\}$ represent the three vertices of the concurrence triangle.
In fact, for $N=3$, from inequality (\ref{4p}) one has $C_{k|ij}(|\psi\rangle)\leq C_{i|jk}(|\psi\rangle)+C_{j|ik}(|\psi\rangle)$ for $i\neq j\neq k\in\{1,2,3\}$. An obvious geometric picture for these inequalities is that the three concurrences represent the lengths of the three edges of a concurrence triangle, see. Fig. 1.
\begin{figure}
  \centering
  \includegraphics[width=10cm]{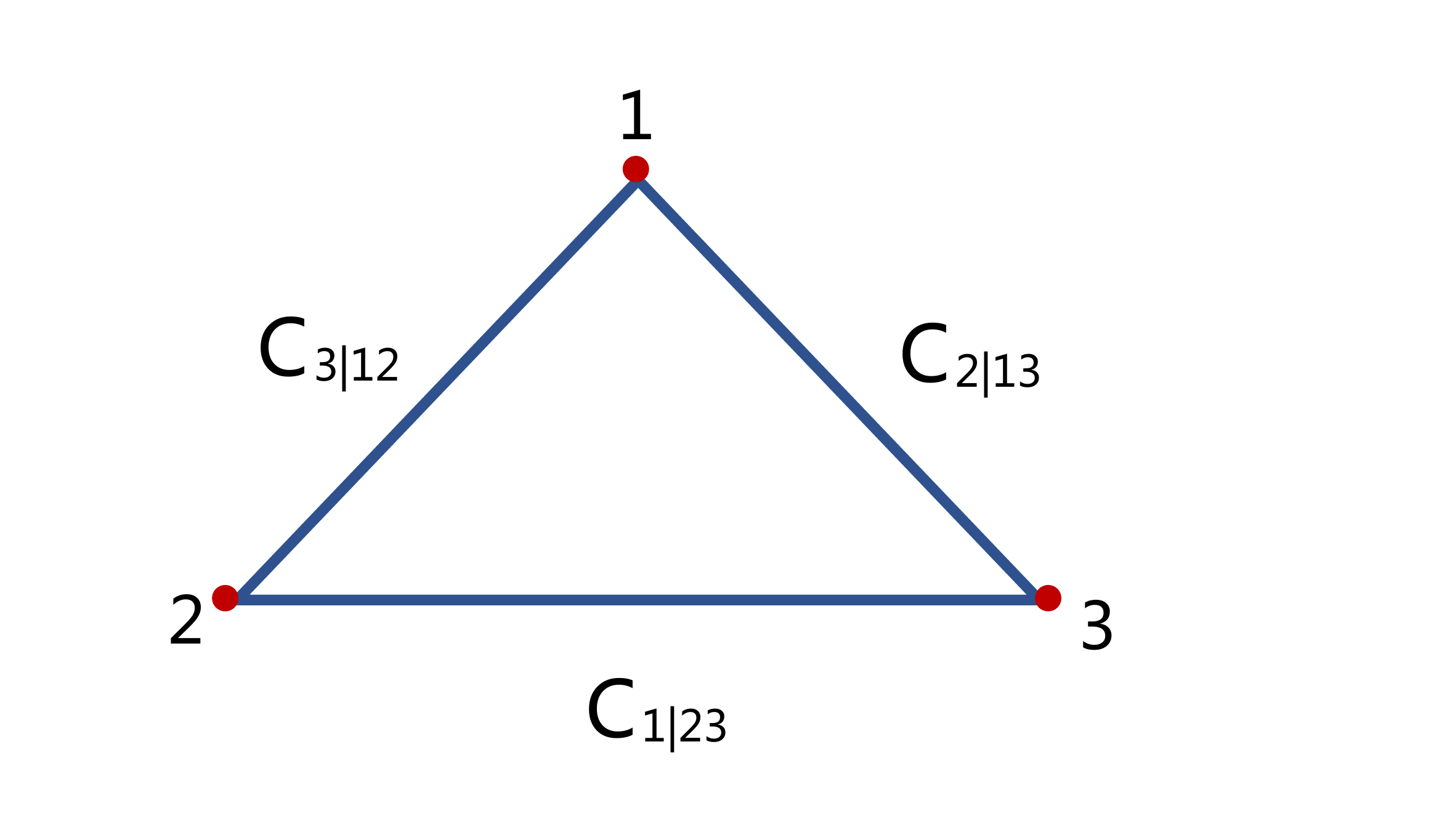}\\
  \caption{The concurrence triangle for a tripartite system. The lengths of the three edges corresponds to the three bipartite concurrences.}
\end{figure}

Set $Q=\frac{1}{2}\sum_{i=1}^3C_{i|\widehat{i}}(|\psi\rangle)$ to be the half-perimeter of the triangle with respect to a tripartite pure state $|\psi\rangle$. We have the following theorem, see proof in section A in Appendix.

{\bf Theorem 2.} For any tripartite pure state $|\psi\rangle$, the area of the concurrence triangle defines a well defined genuine tripartite entanglement measure,
\begin{eqnarray*}\label{prl}
 \mathcal{F}_{3}(|\psi\rangle)=\Big[\frac{16}{3}Q \Pi_{i=1}^3(Q-C_{i|\widehat{i}}(|\psi\rangle)) \Big ]^{\frac{1}{2}},
\end{eqnarray*}
where the factor $\frac{16}{3}$ ensures the normalization $0\leq\mathcal{F}_{3}(|\psi\rangle)\leq 1$.

{\sf Remark 1.} In \cite{xsb}, the authors used squared concurrence as three edges of a triangle, and proposed the following genuine tripartite entanglement measure for three-qubit states,
\begin{eqnarray*}\label{prl}
 F_{123}=\Big[\frac{16}{3}Q \Pi_{i=1}^3(Q-C^2_{i|jk}(|\psi\rangle)) \Big ]^{\frac{1}{4}}.
\end{eqnarray*}
Unfortunately, in \cite{csm} the authors have shown that  $F_{123}$ is increasing under LOCC, which means $F_{123}$ is not a proper genuine entanglement measure. In the following, we give a genuine multipartite entanglement measure based on the geometric mean area of concurrence triangles.

%since the area of a triangle could increase even though the edges are decreased. One can deduce that $F_{123}$ is a proper GME measure under the conditions $C^4_{i|jk}(|\psi\rangle)\leq\sum_{j\ne i}C^4_{j|ik}(|\psi\rangle)$, $i,j,k\in\{1,2,3\}$.

%one can deduce that $F_{123}$ is nonincreasing under LOCC only when $C^4_{i|jk}(|\psi\rangle)\leq\sum_{j\ne i}C^4_{j|ik}(|\psi\rangle),~i,j,k\in\{1,2,3\}$, thus it is not a proper GME measure.

\section{GME Measure for multipartite pure states}
We first consider 4-partite quantum systems. Quite different from the tripartite case, we now need to consider all bipartite entanglement $C_{i| \widehat{i}}$ and $C_{ij|\widehat{ij}}$ for $i,j\in \{1,2,3,4\}$. Thus the polygamy inequalities in (\ref{4p}) for $4$-partite systems correspond to ${P_4^1P_{3}^1}$ concurrence triangles, where $P_n^m=\frac{n!}{m!(n-m)!}$ is the permutation. The corresponding area of the concurrence triangle is given by the Heron's formula, $F_{i|j}=\Big[\frac{16}{3}Q_{i|j} (Q_{i|j}-C_{i| \widehat{i}}(|\psi\rangle)) (Q_{i|j}-C_{j| \widehat{j}}(|\psi\rangle))  (Q_{i|j}-C_{ij|\widehat{ij}}(|\psi\rangle))\Big ]^{\frac{1}{2}}$, where $Q_{i|j}=\frac{1}{2}(C_{i| \widehat{i}}(|\psi\rangle)+ C_{j| \widehat{j}}(|\psi\rangle) +C_{ij|\widehat{ij}}(|\psi\rangle))$ is the half-perimeter and the factor $\frac{16}{3}$ ensures the normalization $0\leq F_{i|j}\leq 1$.
Fig. 2 shows a concurrence triangle for a 4-partite system.
\begin{figure}
  \centering
  \includegraphics[width=8cm]{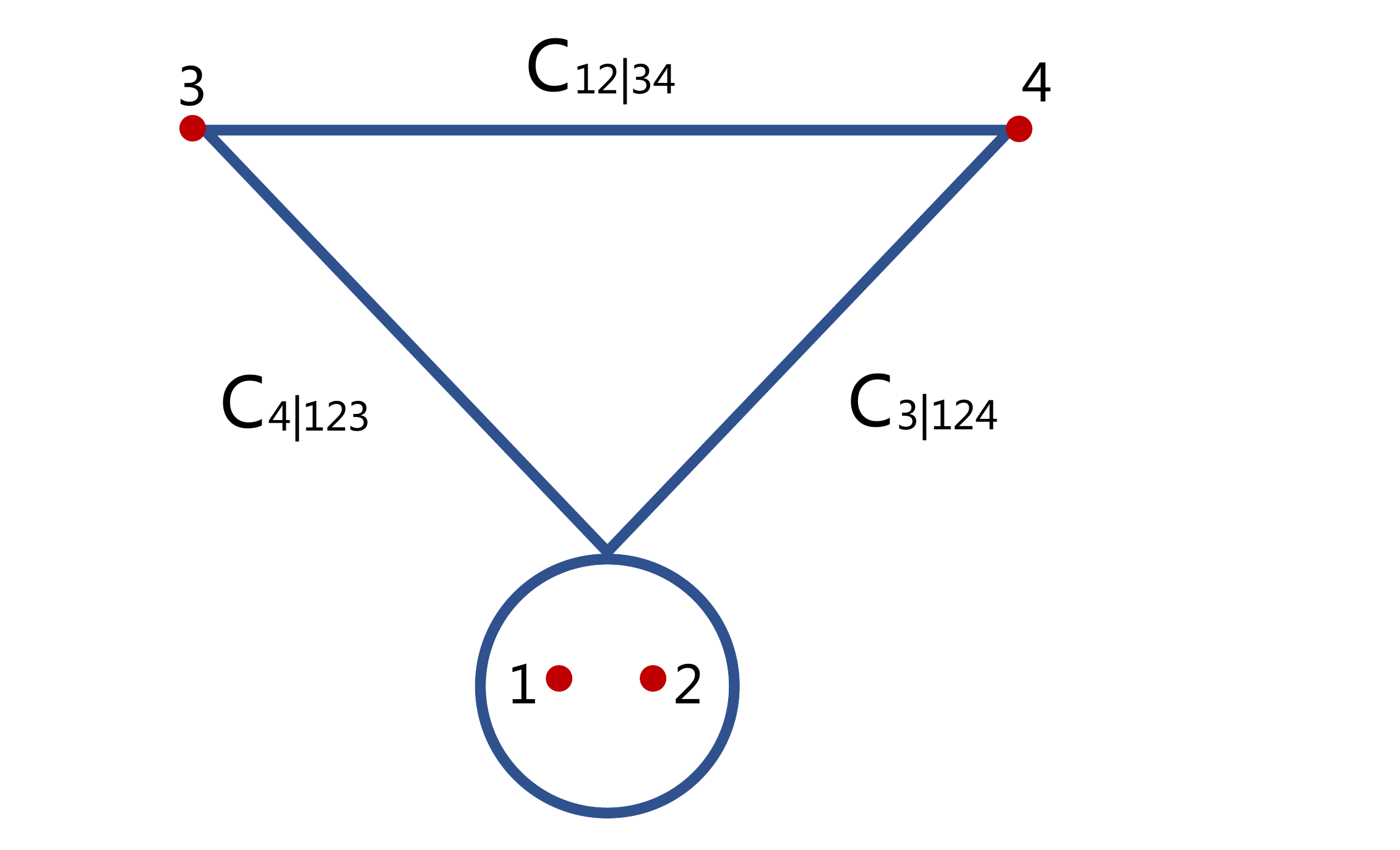}\\
  \caption{A concurrence triangle for a 4-partite system. The lengths of the three edges correspond to the three bipartite concurrences.}
\end{figure}

We use the geometric mean area of the ${P_4^1P_3^1}$ concurrence triangles to define a proper GME measure for 4-partite Pure states. We have the following theorem, see proof in section B in Appendix.

{\bf Theorem 3.} The following geometric mean area of ${P_4^1P_3^1}$ concurrence triangles for 4-partite pure state $|\psi\rangle\in H_{A_1}\otimes ... \otimes H_{A_4}$ is a GME measure,
\begin{eqnarray}\label{PRO}
\mathcal{F}_{4}(|\psi\rangle):=\left(\Pi_{i, j\in\{1,2,3,4\}}F_{i|j}\right)^{\frac{1}{P_4^1P_3^1}}.
\end{eqnarray}

{\sf Remark 2.} Our GME measure (\ref{PRO}) can identify which parts are separable or entangled for non GME states. Actually, a 4-partite pure state $|\psi\rangle$ is not GME if $\mathcal{F}_{4}(|\psi\rangle)=0$. In this case, there must exist some
$i,j\in\{1,2,3,4\}$ such that $F_{i|j}=0$. Without loss of generality, suppose $F_{3|4}=0$, i.e., $(Q_{3|4}-C_{12|34}(|\psi\rangle)) (Q_{3|4}-C_{3| \widehat{3}}(|\psi\rangle)) (Q_{3|4}-C_{4| \widehat{4}}(|\psi\rangle)) =0$. Then $Q_{3|4}$ is equal to at least one of $C_{12|34}(|\psi\rangle)$, $C_{3| \widehat{3}}(|\psi\rangle)$ and $Q_{3|4}=C_{4| \widehat{4}}(|\psi\rangle)$. Assuming $Q_{3|4}=C_{12|34}(|\psi\rangle)$, one gets $C_{3| \widehat{3}}(|\psi\rangle)+ C_{4| \widehat{4}}(|\psi\rangle)=C_{34|12}(|\psi\rangle)$. From the Triangle No-Area Theorem in Ref. \cite{xsb}, at least one of $\{C_{3| \widehat{3}}(|\psi\rangle), C_{4| \widehat{4}}(|\psi\rangle)\}$ is 0. That is to say, $\psi\equiv|\psi\rangle\langle\psi|$ can be expressed by the tensor product of some local density matrices, either $\psi\equiv|\psi\rangle\langle\psi|=\rho_{A_3}\otimes \rho_{A_1A_2A_4} $ or $\rho_{A_4}\otimes \rho_{A_1A_2A_3}$. If $Q_{3|4}= C_{3| \widehat{3}}(|\psi\rangle)$, one gets $C_{3| \widehat{3}}(|\psi\rangle)= C_{4| \widehat{4}}(|\psi\rangle)+C_{12|34}(|\psi\rangle)$. Therefore, $C_{4| \widehat{4}}(|\psi\rangle)=0$ or $C_{12|34}(|\psi\rangle)=0$. This shows that $\psi=\rho_{A_4}\otimes \rho_{A_1A_2A_3}$ or $\psi=\rho_{A_1A_2}\otimes \rho_{A_3A_4}$.
Thus, we identify the separable style of $\psi$.

{\sf Example 1.} For 4-qubit pure states $|GHZ\rangle_{A_1...A_4}=\frac{1}{\sqrt{2}}(|0000\rangle+|1111\rangle) $, we have that $C_{i| \widehat{i}}(|GHZ\rangle)=1$, $C_{ij|\widehat{ij}}(|GHZ\rangle)=1$ and $F_{i|j}(|GHZ\rangle)=1$ for $i,j\in\{1,2,3,4\}$. Then from (\ref{PRO}) we have $\mathcal{F}_{4}(|GHZ\rangle)=1$. For the W state $|W\rangle_{A_1...A_4}=\frac{1}{2}(|1000\rangle+|0100\rangle+|0010\rangle+|0001\rangle)$, we have  $C_{i| \widehat{i}}(|W\rangle)=\frac{3}{4}$, $C_{ij|\widehat{ij}}(|W\rangle)=1$ and $F_{i|j}(|W\rangle)=\left(\frac{5}{12}\right)^\frac{1}{4}$ for $i,j\in\{1,2,3,4\}$. Hence, $\mathcal{F}_{4}(|W\rangle)=\left(\frac{5}{12}\right)^\frac{1}{4}$. Obviously our GME measure (\ref{PRO}) ranks the GHZ state more entangled than the W state \cite{xsb}.

{\sf Example 2.} For the randomly generated 4-qubit pure state $|\psi\rangle_{A_1...A_4}$ with the density matrix $\rho$ given in section C in Appendix, we have $\mathcal{F}_{4}(\psi_{A_1...A_4})=0$, i.e., $|\psi\rangle_{A_1...A_4}$ is not GME. A tedious calculation gives $F_{1|3}=0$, $F_{2|4}=0$ and $C_{34}=0.866$. This implies that $|\psi\rangle_{A_1...A_4}$ can be written as $\psi_{A_1}\otimes \psi_{A_2} \otimes \psi_{A_3A_4}$.  Therefore, there always exist local unitary operators $U_1,U_2,U_3$ and $U_4$ such that $U\rho U^\dagger=|0\rangle_{A_1}|0\rangle_{A_2}(\sin\theta|00\rangle_{A_3A_4}
+\cos\theta|11\rangle_{A_3A_4}) $, where $U=U_1\otimes U_2\otimes U_3\otimes U_4$.

To explore proper GME measures for general $N$-partite pure states, we start with the 5-partite systems. For a 5-partite pure state there are two classifications for the vertices of the concurrence triangles: $\{i,j,\widehat{ij}\}$ and $\{i,jk,\widehat{ijk}\}$, see Fig. 3.
\begin{figure}
  \centering
  \includegraphics[width=8cm]{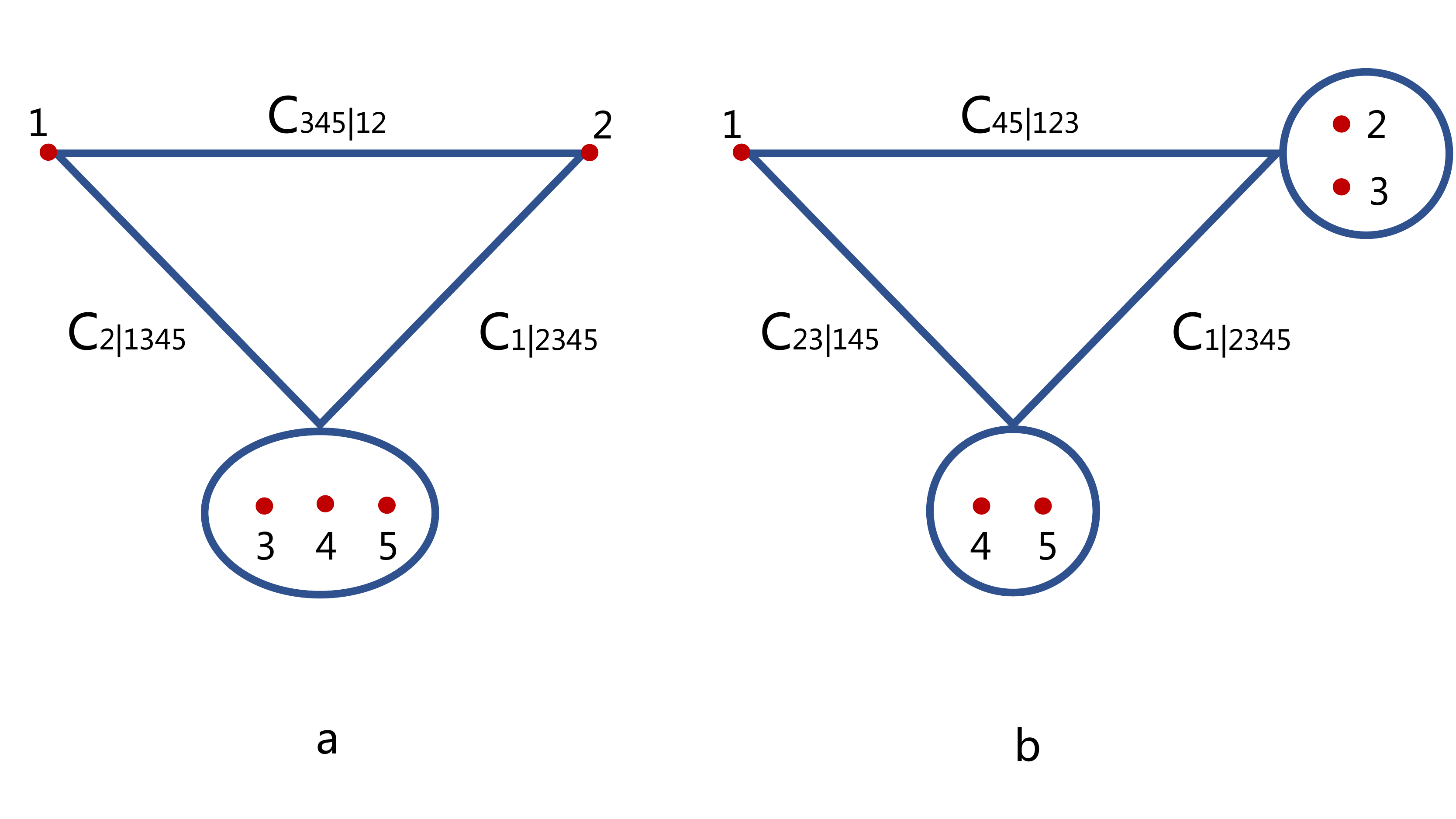}\\
  \caption{The concurrence triangle for a 5-partite pure states. Fig. a corresponds to $\mathcal{F}_{5}^{(1)}(|\psi\rangle)$ with subsystems $3,4,5$ as a group. Fig. b corresponds to $\mathcal{F}_{5}^{(2)}(|\psi\rangle)$ with subsystems $2,3$ and $4,5$ as groups, respectively. The lengths of the three edges correspond to the three bipartite concurrences.}
\end{figure}

For the case $\{i,j,\widehat{ij}\}$, using the result in Theorem 3, we have a geometric mean area of $P_5^1P_4^1$ concurrence triangles,
\begin{eqnarray*}\label{5part1}
\mathcal{F}_{5}^{(1)}(|\psi\rangle):=\left(\Pi_{i, j \in\{1,2,3,4,5\}}F_{i|j}\right)^{\frac{1}{P_5^1P_4^1}},
\end{eqnarray*}
where each $F_{i|j}$ is the area of a concurrence triangle whose three edges are $C_{i| \widehat{i}}(|\psi\rangle), C_{j| \widehat{j}}(|\psi\rangle)$ and $C_{ij|\widehat{ij}}(|\psi\rangle)$, respectively.

For the case $\{i,jk,\widehat{ijk}\}$, similar to Theorem 3, one has another geometric mean area of $P_5^1P_4^2$ concurrence triangles,
\begin{eqnarray*}\label{5part2}
\mathcal{F}_{5}^{(2)}(|\psi\rangle):=\left(\Pi_{i, jk\in\{1,2,3,4,5\}}F_{i|\vec{j}_2}\right)^{\frac{1}{P_5^1P_4^2}},
\end{eqnarray*}
where each $F_{i|jk}$ is the area of a concurrence triangle whose three edges are $C_{i| \widehat{i}}(|\psi\rangle)$, $C_{jk| \widehat{jk}}(|\psi\rangle)$ and $C_{\widehat{ijk}|ijk}(|\psi\rangle)$.

Although the expressions of $\mathcal{F}_{5}^{(1)}(|\psi\rangle)$ and $\mathcal{F}_{5}^{(2)}(|\psi\rangle)$ may be different, these two GME measures $\mathcal{F}_{5}^{(1)}(|\psi\rangle)$ and $\mathcal{F}_{5}^{(2)}(|\psi\rangle)$ are equivalent in the sense that $\mathcal{F}_{5}^{(1)}(|\psi\rangle)>0$ if and only if $\mathcal{F}_{5}^{(2)}(|\psi\rangle)>0$.
Actually, $\mathcal{F}_{5}^{(1)}(|\psi\rangle)=0$ iff each $F_{i|j}=0$, $i,j \in\{1,2,3,4,5\}$, iff at least one of the three edges $C_{i| \widehat{i}}(|\psi\rangle)$, $C_{j| \widehat{j}}(|\psi\rangle)$ and $C_{\widehat{ij}|ij}(|\psi\rangle)$ is zero. Similarly,
$\mathcal{F}_{5}^{(2)}(|\psi\rangle)=0$ iff $F_{i|\vec{j}_2}=0$, $i, \vec{j}_2\in\{1,2,3,4,5\}$, iff at least one of the three edges $C_{i| \widehat{i}}(|\psi\rangle)$, $C_{\vec{j}_2| \widehat{\vec{j}_2}}(|\psi\rangle)$ and $C_{\widehat{i\vec{j}_2}|i\vec{j}_2}(|\psi\rangle)$ is zero.
Since $C_{\vec{j}_2| \widehat{\vec{j}_2}}(|\psi\rangle)=C_{\widehat{ij}|ij}(|\psi\rangle)$ and the values of $C_{\widehat{i\vec{j}_2}|i\vec{j}_2}(|\psi\rangle)$ one to one correspond to that of $C_{\widehat{ij}|ij}(|\psi\rangle)$, we obtain $\mathcal{F}_{5}^{(1)}(|\psi\rangle)=0$ iff $\mathcal{F}_{5}^{(2)}(|\psi\rangle)=0$. Thus we have two different but equivalent GME measures for 5-partite pure states.

We are ready now to define the GME measures for general multipartite systems. We present a proper GME measure for $N$-partite pure state as follows, see proof in section D in Appendix.

{\bf Theorem 4.} The geometric mean area of $\mathcal{F}_{N}^{(l)}(|\psi\rangle)$, $1\leq l\leq [\frac{N-2}{2}]$, is a proper GME measure for any $N$-partite pure state $|\psi\rangle\in H_{A_1}\otimes...\otimes H_{A_N}$,
\begin{eqnarray}\label{th3}
\mathcal{F}_{N}(|\psi\rangle):=\left(\Pi_{l=1} ^{[\frac{N-2}{2}]}\mathcal{F}_{N}^{(l)}(|\psi\rangle)\right)^\frac{1}{[\frac{N-2}{2}]},
\end{eqnarray}
where $[\cdot]$ stands for rounding down.

{\sf Remark 3.} $\mathcal{F}_{N}(|\psi\rangle)$ in Eq. (\ref{th3}) is just a representative of GME measures. From different classifications of the concurrence triangles, other GME measures can be obtained, which are essentially equivalent to each other, similar to the case of  5-partite systems.

\section{GME measures for mixed states}
The geometric mean area of the concurrence triangles can also be conceptually generalized to multipartite mixed states via the convex roof construction:
\begin{eqnarray}\label{mix}
\mathcal{F}_N(\rho):=\min_{\{p_i |\psi_i\rangle\}}p_i \mathcal{F}_N(|\psi_i\rangle),
\end{eqnarray}
where $\rho$ is an $N$-partite mixed state, and the minimum is taken over all possible pure state decompositions $\rho=\sum_ip_i |\psi_i\rangle$.

Obviously, $\mathcal{F}_N(\rho)>0$ implies at least one $\mathcal{F}_N(|\psi_i\rangle)>0$, i.e., $\rho$ is a GME state. While $\mathcal{F}_N(\rho)=0$ implies each $\mathcal{F}_N(\psi_i)=0$, i.e., every pure state is at least bi-separable, thus $\rho$ is not GME. From Eq.(\ref{th3}), for any LOCC operator $\Lambda$, $\mathcal{F}_N(|\psi_i\rangle)$ is nonincresing under LOCC, i.e., $\mathcal{F}_N(\Lambda(|\psi_i\rangle)) \leq \mathcal{F}_N(|\psi_i\rangle)$. Thus, suppose $\rho=\sum_{i'}p_{i'} \psi_{i'}$ is an optimal decomposition in (\ref{mix}). Then $\mathcal{F}_N(\Lambda(\rho))=\sum_{{i'}}p_{i'} \mathcal{F}_N(\Lambda(|\psi_{i'}\rangle))\leq \sum_{{i'}}p_{i'} \mathcal{F}_N(|\psi_{i'}\rangle)=\mathcal{F}_N(\rho)$, i.e., $\mathcal{F}_N(\rho)$ is nonincresing under LOCC. Therefore, the GME measure defined by Eq. (\ref{mix}) for mixed states is a proper one.

In general, it is hard to calculate $\mathcal{F}_N(\rho)$ due to infinite many pure state ensemble decompositions. In the following, we give a witness to determine whether a mixed state is GME or not. For a general $N$-partite mixed state $\rho_A:=\rho_{A_1...A_N}=\sum_{i} p_i|i\rangle_A\langle i|$, there exists a reference system $R$ such that  $|\psi\rangle_{AR}=\sum_{i}\sqrt{p_i}|i\rangle_A|i\rangle_R$ is the purified state of $\rho_A$, where $|i\rangle_A$ and $|i\rangle_R$ are the bases of subsystems $A$ and $R$, respectively. We can define a GME witness for mixed states as follows,
\begin{eqnarray}\label{mixed}
\mathcal{F}_{N}(\rho_A):=\mathcal{F}_{N+1}(|\psi\rangle_{AR}).
\end{eqnarray}
If $\mathcal{F}_{N}(\rho_A)=0$, then $\rho_A$ is not a GME state. If $\rho_A$ is GME, then one gets $\mathcal{F}_{N}(\rho_A)>0$.
%For example, suppose $F_{i|j}=0$ for some $i,j\in\{1,2,\cdots N\}$, one gets $\psi_{AR}=\rho_{A_{i(j)}}\otimes \rho_{A_{\widehat{i(j)}R}}$ or $\psi_{AR}=\rho_{A_iA_j}\otimes \rho_{A_{\widehat{ij}R}}$. In other words, $\rho_A$ can be expressed as $\rho_A=\rho_{A_{i(j)}}\otimes \rho'$ or $\rho_A=\rho_{A_iA_j}\otimes \rho''$, where $\rho'=\mathrm{Tr}_R\rho_{A_{\widehat{i(j)}R}}$ and $\rho''=\mathrm{Tr}_R\rho_{A_{\widehat{ij}R}}$ are mixed states.

Let us consider the 4-qubit pure state $|\psi\rangle_{A_1...A_4}$ in Example 2. Denote $\rho'=\Tr_{A_3}|\psi\rangle_{A_1...A_4}\langle\psi|$. $\rho'$ is a generally a mixed state. Combining the results in Example 2 with (\ref{mixed}) and taking $R=A_3$ as the reference system, we have $\mathcal{F}_{3}(\rho')=\mathcal{F}_{4}(|\psi\rangle_{A_1...A_4})=0$. As $|\psi\rangle_{A_1...A_4}$ is local unitary equivalent to the state $|0\rangle_{A_1}|0\rangle_{A_2}(\sin\theta|00\rangle_{A_3A_4}+\cos\theta|11\rangle_{A_3A_4})$, we get that $\rho'$ is not a GME. 

{\sf Example 3.} For the three-qubit mixed state $\rho_{A_1A_2A_3}$ given in section E in Appendix, direct calculation shows that $C(\rho_{A_iA_j})=\frac{1}{2}$, $\forall i,j \in\{1,2,3\}$, which verifies that $\rho_{A_1A_2A_3}$ is GME.
The eigenvalues of $\rho_{A_1A_2A_3}$ are $\frac{3}{4}$ and $\frac{1}{4}$, with the corresponding eigenvectors $|\alpha_0\rangle$ and $|\alpha_1\rangle$, respectively. We can add a qubit system $R$ with basis $\{|0\rangle,|1\rangle\}$ such that $|\psi\rangle_{A_1A_2A_3R}=\frac{\sqrt{3}}{2}|\alpha_0\rangle|0\rangle
+\frac{1}{2}|\alpha_1\rangle|1\rangle$. Then from Eq. (\ref{th3}) and Eq. (\ref{mixed}), we have $\mathcal{F}_{3}(\rho_{A_1A_2A_3})=\mathcal{F}_{4}(\psi_{A_1A_2A_3R})=0.8034>0$.

\section{Conclusion}
Quantum multipartite entanglement plays an important role in quantum information theory.
Proper GME measures to quantify the genuine multipartite entanglement faithfully are of great significance. We have presented a set of polygamy inequalities satisfied by multipartite pure states, with geometric interpretations in terms of concurrence triangles.
For any $N$-partite pure states, we have advanced a bona fide GME measure based on the geometric mean area of concurrence triangles. Furthermore, if a pure state is not GME, we can certify which part is separable with the remaining by our GME measure. The GME measure of multipartite mixed states has been obtained via the convex roof construction. Based on state purification, we have also presented a witness which detects the GME of general mixed multipartite entanglement. Our approach may also be used to study the ``genuine properties" related to other quantum correlations such as genuine nonlocality.

\medskip

{\it Acknowledgments.} This work is supported by the National Natural Science Foundation of China (NSFC) under Grants 11847209, 11761073, 12075159 12171044 and 61727801; Beijing Natural Science Foundation (Grant No. Z190005); the China Postdoctoral Science Foundation funded project No. 2019M650811 and the China Scholarship Council No. 201904910005; Academician Innovation Platform of Hainan Province.

\medskip

{\it Note added.} After completing this work, it is found that Guo et al. \cite{guoy} also studied this problem by using the product of the radius of inscribed circle and circumscribed circle of the concurrence triangle. The method they used and the constructed GME measure are completely different from ours.

\bigskip
\section*{APPENDIX}

\maketitle

\section*{A. Proof of Theorem 2}
We first prove that $\mathcal{F}_{3}(|\psi\rangle)$ is GME iff $\mathcal{F}_{3}(|\psi\rangle) >0$. On one hand, if $\mathcal{F}_{3}(|\psi\rangle)>0$, then each edge of the concurrence triangle is positive, that is to say, $C_{i|jk}(|\psi\rangle)$, $C_{j|ik}(|\psi\rangle)$ and $C_{k|ij}(|\psi\rangle)$ are all positive. Hence $|\psi\rangle$ is GME. On the other hand, if $\mathcal{F}_{3}(|\psi\rangle)=0$, then from the area of the concurrence triangle is zero iff the length of at least one edge is zero \cite{xsb}. We obtain that at least one of
$\{C_{1|23}(|\psi\rangle)$, $C_{2|13}(|\psi\rangle)$ and $C_{3|12}(|\psi\rangle)\}$ is 0, namely, either $\psi=\rho_{A_1}\otimes \rho_{A_2A_3} $ or $\psi=\rho_{A_2}\otimes \rho_{A_1A_3}$ or $\psi=\rho_{A_3}\otimes \rho_{A_1A_2}$. Therefore, $\psi$ is not GME.

We next prove that $\mathcal{F}_{3}(|\psi\rangle)$ cannot increase under LOCC, i.e., $\mathcal{F}_{3}(|\psi\rangle)\geq \mathcal{F}_{3}(\Lambda(|\psi\rangle))$ for any LOCC map $\Lambda$. As the concurrence $C$ is an entanglement measure, it is nonincreasing under LOCC. Hence we only need to prove that $\mathcal{F}_{3}(|\psi\rangle)$ is an increasing function of $C_{i|jk}(|\psi\rangle)$, $C_{j|ik}(|\psi\rangle)$ and $C_{k|ij}(|\psi\rangle)$. Since the monotonicity of $\mathcal{F}_{3}(|\psi\rangle)$ is in consistent with the monotonicity of its square, denoting by $\mathcal{G}_{3}(|\psi\rangle):=Q \Pi_{i=1}^3(Q-C_{i|jk}(|\psi\rangle))$, we consider the derivative of $\mathcal{G}_{3}(|\psi\rangle)$ with respect to $C_{i|jk}(|\psi\rangle)$, $\frac{\partial{\mathcal{G}}_{3}(|\psi\rangle)}{\partial{C}_{i|jk}(|\psi\rangle)}
=\frac{1}{4}C_{i|jk}(|\psi\rangle)(C^2_{j|ik}(|\psi\rangle)+ C^2_{k|ij}(|\psi\rangle) -C^2_{i|jk}(|\psi\rangle))$. From Eq. (\ref{th1}), we have $C^2_{j|ik}(|\psi\rangle)+ C^2_{k|ij}(|\psi\rangle) \geq C^2_{i|jk}(|\psi\rangle)$. One gets
$\frac{\partial{\mathcal{G}}_{3}(|\psi\rangle)}{\partial{C}_{i|jk}(|\psi\rangle)}\geq 0$. Similarly, we have $\frac{\partial{\mathcal{G}}_{3}(|\psi\rangle)}{\partial{C}_{j|ik}(|\psi\rangle)}\geq 0$ and $\frac{\partial{\mathcal{G}}_{3}(|\psi\rangle)}{\partial{C}_{k|ij}(|\psi\rangle)}\geq 0$. Thus the monotonicity of $\mathcal{F}_{3}(|\psi\rangle)$ holds and hence $\mathcal{F}_{3}(|\psi\rangle)$ is non-increasing under LOCC.

\section*{B. Proof of Theorem 3}

We first prove that $\mathcal{F}_{4}(|\psi\rangle)$ is GME iff $ \mathcal{F}_{4}(|\psi\rangle) >0$. On one hand, if $\mathcal{F}_{4}(|\psi\rangle)>0$, then each triangle area $F_{i|j}>0$,  $i,j\in\{1,2,3,4\}$. Hence each edge of the ${P_4^1P_3^1}$ concurrence triangles is positive, i.e., $C_{i| \widehat{i}}(|\psi\rangle)$, $C_{j| \widehat{j}}(|\psi\rangle)$ and $C_{\widehat{ij}|ij}(|\psi\rangle)$ are all positive. Therefore, $\psi$ is GME. On the other hand, if $\mathcal{F}_{4}(|\psi\rangle)=0$, then $F_{i|j}=0$ for some $i,j\in\{1,2,3,4\}$. Without loss of generality, suppose $F_{3|4}=0$. As the area of the concurrence triangle is zero iff it has at least one edge with zero length \cite{xsb}, we obtain that at least one of
$\{C_{3| \widehat{3}}(|\psi\rangle)$, $C_{4| \widehat{4}}(|\psi\rangle)$ and $C_{12|34}(|\psi\rangle)\}$ is 0, that is to say, either $\psi=\rho_{A_3}\otimes \rho_{A_1A_2A_4} $ or $\psi=\rho_{A_4}\otimes \rho_{A_1A_2A_3}$ or $\psi=\rho_{A_1A_2}\otimes \rho_{A_3A_4}$. Therefore, $\psi$ is not GME.

We next to prove that $\mathcal{F}_{4}(|\psi\rangle)$ does not increase under LOCC. Obviously we only need to verify that each $F_{i|j}$ cannot increase under LOCC, i.e., $F_{i|j}(|\psi\rangle)\geq F_{i|j}(\Lambda(|\psi\rangle))$ for any LOCC map $\Lambda$. As the concurrence is non-increasing under LOCC, we only need to prove that $F_{i|j}(|\psi\rangle)$ is an increasing function of $C_{i| \widehat{i}}(|\psi\rangle)$, $C_{j| \widehat{j}}(|\psi\rangle)$ and $C_{\widehat{ij}|ij}(|\psi\rangle)$. Since the monotonicity of $F_{i|j}(|\psi\rangle)$ is in consistent with the monotonicity of its square, denoting by $G_{i|j}(|\psi\rangle):=Q_{i|j} (Q_{i|j}-C_{i| \widehat{i}}(|\psi\rangle)) (Q_{i|j}-C_{j| \widehat{j}}(|\psi\rangle))  (Q_{i|j}-C_{\widehat{ij}|ij}(|\psi\rangle))$, we consider the derivative, $\frac{\partial{G}_{i|j}(|\psi\rangle)}{\partial{C}_{i| \widehat{i}}(|\psi\rangle)}=\frac{1}{4}C_{i| \widehat{i}}(|\psi\rangle)(C^2_{j| \widehat{j}}(|\psi\rangle)+ C^2_{\widehat{ij}|ij}(|\psi\rangle) -C^2_{i| \widehat{i}}(|\psi\rangle))$. From Eq. (\ref{th1}), we have $C^2_{j| \widehat{j}}(|\psi\rangle)+ C^2_{\widehat{ij}|ij}(|\psi\rangle) \geq C^2_{i| \widehat{i}}(|\psi\rangle)$. Hence we have $\frac{\partial{G}_{i|j}(|\psi\rangle)}{\partial{C}_{i| \widehat{i}}(|\psi\rangle)}\geq 0$. Similarly, we can obtain $\frac{\partial{G}_{i|j}(|\psi\rangle)}{\partial{C}_{j| \widehat{j}}(|\psi\rangle)}\geq 0$ and $\frac{\partial{G}_{i|j}(|\psi\rangle)}{\partial{C}_{\widehat{ij}|ij}(|\psi\rangle)}\geq 0$. Thus the monotonicity of $F_{i|j}(|\psi\rangle)$ holds and hence $\mathcal{F}_{4}(|\psi\rangle)$ is non-increasing under LOCC.

\section*{C. Example 2.}

\begin{equation*}       %开始数学环境
\rho =\left(                 %左括号
  \begin{array}{cccc}   %该矩阵一共3列，每一列都居中放置
\rho_{11} &\rho_{12} & \rho_{13} & \rho_{14}\\  %第一行元素
\rho^\dagger_{12}& \rho_{22} & \rho_{23}& \rho_{24}\\  %第二行元素
\rho^\dagger_{13}&\rho^\dagger_{23}& \rho_{33}&\rho_{34}\\
\rho^\dagger_{14}& \rho^\dagger_{24}&\rho^\dagger_{34}& \rho_{44}\\
  \end{array}
\right),                 %右括号
\end{equation*}
where $\rho^\dagger$ is the conjugate transpose of $\rho$ and
\begin{equation*}       %开始数学环境
\rho_{11}=\left(                 %左括号
  \begin{array}{cccc}   %该矩阵一共3列，每一列都居中放置
0.0247608 & 0.0279231 & -0.0279231 & 0.00748196\\  %第一行元素
0.0279231& 0.0314892 & -0.0314892& 0.0084375\\  %第二行元素
-0.0279231&-0.0314892& 0.0314892& -0.0084375\\
0.00748196& 0.0084375& -0.0084375& 0.00226082\\
  \end{array}
\right)                 %右括号
\end{equation*}

\begin{equation*}       %开始数学环境
\rho_{12}=\left(                 %左括号
  \begin{array}{cccc}   %该矩阵一共3列，每一列都居中放置
0.0330144 & 0.0372308& -0.0372308& 0.00997595\\  %第一行元素
0.0372308& 0.0419856& -0.0419856& 0.01125\\  %第二行元素
-0.0372308& -0.0419856& 0.0419856& -0.01125\\
0.00997595& 0.01125& -0.01125& 0.00301443\\
  \end{array}
\right)                 %右括号
\end{equation*}

\begin{equation*}       %开始数学环境
\rho_{13}=\left(                 %左括号
  \begin{array}{cccc}   %该矩阵一共3列，每一列都居中放置
0.042887& 0.0483642& -0.0483642& 0.0129591\\  %第一行元素
0.0483642& 0.0545409& -0.0545409& 0.0146142\\  %第二行元素
-0.0483642& -0.0545409& 0.0545409& -0.0146142\\
0.0129591&0.0146142& -0.0146142& 0.00391586\\
  \end{array}
\right)                 %右括号
\end{equation*}

\begin{equation*}       %开始数学环境
\rho_{14}=\left(                 %左括号
  \begin{array}{cccc}   %该矩阵一共3列，每一列都居中放置
0.0571827& 0.0644856& -0.0644856& 0.0172789\\  %第一行元素
0.0644856& 0.0727211& -0.0727211& 0.0194856\\  %第二行元素
-0.0644856& -0.0727211& 0.0727211& -0.0194856\\
0.0172789& 0.0194856& -0.0194856& 0.00522114\\
  \end{array}
\right)                 %右括号
\end{equation*}

\begin{equation*}       %开始数学环境
\rho_{22}=\left(                 %左括号
  \begin{array}{cccc}   %该矩阵一共3列，每一列都居中放置
0.0440192& 0.049641& -0.049641& 0.0133013\\
0.049641& 0.0559808& -0.0559808& 0.015\\
-0.049641& -0.0559808& 0.0559808& -0.015\\
0.0133013& 0.015& -0.015& 0.00401924\\
  \end{array}
\right)                 %右括号
\end{equation*}

\begin{equation*}       %开始数学环境
\rho_{23}=\left(                 %左括号
  \begin{array}{cccc}   %该矩阵一共3列，每一列都居中放置
0.0571827& 0.0644856& -0.0644856& 0.0172789\\
0.0644856& 0.0727211& -0.0727211& 0.0194856\\
-0.0644856& -0.0727211& 0.0727211& -0.0194856\\
0.0172789& 0.0194856& -0.0194856& 0.00522114\\
  \end{array}
\right)                 %右括号
\end{equation*}

\begin{equation*}       %开始数学环境
\rho_{24}=\left(                 %左括号
  \begin{array}{cccc}   %该矩阵一共3列，每一列都居中放置
0.0762436& 0.0859808& -0.0859808& 0.0230385\\
0.0859808& 0.0969615& -0.0969615& 0.0259808\\
-0.0859808& -0.0969615& 0.0969615& -0.0259808\\
0.0230385& 0.0259808& -0.0259808& 0.00696152\\
  \end{array}
\right)                 %右括号
\end{equation*}

\begin{equation*}       %开始数学环境
\rho_{33}=\left(                 %左括号
  \begin{array}{cccc}   %该矩阵一共3列，每一列都居中放置
0.0742825& 0.0837692& -0.0837692& 0.0224459\\
0.0837692& 0.0944675& -0.0944675& 0.0253125\\
-0.0837692& -0.0944675& 0.0944675& -0.0253125\\
0.0224459& 0.0253125& -0.0253125& 0.00678246 \\
  \end{array}
\right)                 %右括号
\end{equation*}

\begin{equation*}       %开始数学环境
\rho_{34}=\left(                 %左括号
  \begin{array}{cccc}   %该矩阵一共3列，每一列都居中放置
0.0990433& 0.111692& -0.111692& 0.0299279\\
0.111692& 0.125957& -0.125957& 0.03375\\
-0.111692& -0.125957& 0.125957& -0.03375\\
0.0299279& 0.03375& -0.03375& 0.00904329\\
  \end{array}
\right)                 %右括号
\end{equation*}

\begin{equation*}       %开始数学环境
\rho_{44}=\left(                 %左括号
  \begin{array}{cccc}   %该矩阵一共3列，每一列都居中放置
0.132058& 0.148923& -0.148923& 0.0399038\\
0.148923& 0.167942& -0.167942& 0.045\\
-0.148923& -0.167942& 0.167942& -0.045\\
0.0399038& 0.045& -0.045& 0.0120577\\
  \end{array}
\right)                 %右括号
\end{equation*}

\section*{D. Proof of Theorem 4.}
By induction, for $N$-partite pure states, we only need to consider concurrence triangles for the following $[\frac{N-2}{2}]$ cases, where $[\cdot]$ means rounding down. Denote $\vec{j}_m=(j_1j_2\cdots j_m)$ for convenience.

Case 1). The three vertices are $\{i,j,\widehat{ij}\}$ and the geometric mean area of the concurrence triangles is
\begin{eqnarray*}\label{Npart1}
\mathcal{F}_{N}^{(1)}(|\psi\rangle):=\left(\Pi_{i, j\in\{1,2,\cdots, N\}}F_{i|j}\right)^{\frac{1}{P_N^1P_{N-1}^1}}.
\end{eqnarray*}

Case 2). The three vertices are $\{i,\vec{j}_2, \widehat{i\vec{j}}_2\}$ and the geometric mean area of concurrence triangles is
\begin{eqnarray*}\label{Npart2}
\mathcal{F}_{N}^{(2)}(|\psi\rangle):=\left(\Pi_{i, \vec{j}_2\in\{1,2,\cdots, N\}}F_{i|\vec{j}_2}\right)^{\frac{1}{P_N^1P_{N-1}^2}},
\end{eqnarray*}

$$......$$

Case $[\frac{N-2}{2}]).$ The three vertices are $\{i, \vec{j}_{[\frac{N-2}{2}]}, \widehat{i\vec{j}}_{[\frac{N-2}{2}]}\}$ and the geometric mean area of concurrence triangles is
\begin{eqnarray*}\label{Npart2}
\mathcal{F}_{N}^{([\frac{N-2}{2}])}(|\psi\rangle):=\left(\Pi_{i, \vec{j}_{[\frac{N-2}{2}]}\in\{1,2,\cdots, N\}}F_{i|\vec{j}_{[\frac{N-2}{2}]}}\right)^\frac{1}{P_N^1P_{N-1}^{[\frac{N-2}{2}]}},
\end{eqnarray*}
where each $F_{i|\vec{j}_{[\frac{N-2}{2}]}}$ is the area of the concurrence triangle whose edges are $C_{i| \widehat{i}}(|\psi\rangle)$, $C_{\vec{j}_{[\frac{N-2}{2}]}| \widehat{\vec{j}}_{[\frac{N-2}{2}]}}(|\psi\rangle)$ and $C_{\widehat{i\vec{j}}_{[\frac{N-2}{2}]}|i\vec{j}_{[\frac{N-2}{2}]}}(|\psi\rangle)$.

The reason why we only need to consider these $[\frac{N-2}{2}]$ cases for $N$-partite pure states is that $\mathcal{F}_{N}^{(n)}(|\psi\rangle)$ $([\frac{N-2}{2}]<n<N-1)$ are equivalent to $\mathcal{F}_{N}^{(m)}(|\psi\rangle)$ $(1\leq m\leq [\frac{N-2}{2}])$, similar to the equivalence between $\mathcal{F}_{5}^{(1)}(|\psi\rangle)$ and $\mathcal{F}_{5}^{(2)}(|\psi\rangle)$ in 5-partite systems.

Every case corresponds to a geometric mean area of concurrence triangles, and $\mathcal{F}_{N}^{(l)}(|\psi\rangle)>0$, $l=1,2,...,[\frac{N-2}{2}]$ iff the state is entangled under any bipartition. Therefore, the geometric mean of $[\frac{N-2}{2}]$ cases defines a well-defined GME measure for $N$-partite states.

\section*{E. Example 3.}

\begin{widetext}
\begin{equation*}       %开始数学环境
\rho_{A_1A_2A_3}=\left(                 %左括号
  \begin{array}{cccccccc}   %该矩阵一共3列，每一列都居中放置
0.15625& -0.241627& -0.133373& -0.09375&0.270633& -0.41851& -0.23101& -0.16238\\
-0.241627& 0.560256& 0.15625& 0.00837341& -0.41851& 0.970392& 0.270633& 0.0145032\\
-0.133373& 0.15625& 0.127244& 0.116627& -0.23101& 0.270633& 0.220392& 0.202003\\
-0.09375& 0.00837341& 0.116627& 0.15625& -0.16238& 0.0145032&  0.202003& 0.270633\\
0.270633& -0.41851& -0.23101& -0.16238& 0.46875& -0.72488& -0.40012& -0.28125\\
-0.41851& 0.970392& 0.270633& 0.0145032& -0.72488& 1.68077& 0.46875& 0.0251202\\
-0.23101& 0.270633& 0.220392& 0.202003& -0.40012& 0.46875& 0.381731& 0.34988\\
-0.16238& 0.0145032& 0.202003& 0.270633& -0.28125& 0.0251202& 0.34988& 0.46875\\
  \end{array}
\right)                 %右括号
\end{equation*}
\end{widetext}

\end{document}